# Supervised machine learning based multi-task artificial intelligence classification of retinopathies

**Minhaj Alam[1], David Le[1], Jennifer I. Lim[2], R.V.P. Chan[2], and Xincheng Yao[1,2]**

[1]*Department of Bioengineering, University of Illinois at Chicago, Chicago, IL 60607, USA*

[2]*Department of Ophthalmology and Visual Sciences, University of Illinois at Chicago, Chicago, IL 60612, USA*

**Running head:** Multi-task AI classification of retinopathies

**Corresponding author:**

Xincheng Yao, PhD

Richard & Loan Hill Professor, Department of Bioengineering (MC 563)

Professor, Department of Ophthalmology & Visual Sciences

University of Illinois at Chicago (UIC)

Clinical Sciences North, Suite W103, Room 164D

820 South Wood Street, Chicago, IL 60612

Tel: (312)413-2016; Fax: (312)996-4644

Email: xcy@uic.edu.

**Financial support:** This research was supported in part by NIH grants R01 EY030101, R01 EY024628, P30 EY001792; by unrestricted grant from Research to Prevent Blindness; by Richard and Loan Hill endowment; by Marion H. Schenk Chair endowment.

**Competing Interest:** Pending patent application: X. Yao, M. Alam, and J. I. Lim No competing interest exists for any other author.

**Acknowledgement:** The authors thank Mr. Mark Janowicz and Ms. Andrea Degillio (Eye and Ear Infirmary, University of Illinois at Chicago) for technical support of data acquisition.

# Abstract

Artificial intelligence (AI) classification holds promise as a novel and affordable screening tool for clinical management of ocular diseases. Rural and underserved areas, which suffer from lack of access to experienced ophthalmologists may particularly benefit from this technology. Quantitative optical coherence tomography angiography (OCTA) imaging provides excellent capability to identify subtle vascular distortions, which are useful for classifying retinovascular diseases. However, application of AI for differentiation and classification of multiple eye diseases is not yet established. In this study, we demonstrate supervised machine learning based multi-task OCTA classification. We sought 1) to differentiate normal from diseased ocular conditions, 2) to differentiate different ocular disease conditions from each other, and 3) to stage the severity of each ocular condition. Quantitative OCTA features, including blood vessel tortuosity (BVT), blood vascular caliber (BVC), vessel perimeter index (VPI), blood vessel density (BVD), foveal avascular zone (FAZ) area (FAZ-A), and FAZ contour irregularity (FAZ-CI) were fully automatically extracted from the OCTA images. A stepwise backward elimination approach was employed to identify sensitive OCTA features and optimal-feature-combinations for the multi-task classification. For proof-of-concept demonstration, diabetic retinopathy (DR) and sickle cell retinopathy (SCR) were used to validate the supervised machine leaning classifier. The presented AI classification methodology is applicable and can be readily extended to other ocular diseases, holding promise to enable a mass-screening platform for clinical deployment and telemedicine.

Machine learning based artificial intelligence (AI) technology has garnered increasing interest in medical applications over the past few years.[1] An AI-based diagnosis tool is designed to mimic the perception of the human brain for information processing and making objective decisions. Recent studies have demonstrated AI applications in detecting retinal disease progression,[2-5] identifying malignant or benign melanoma,[6] and classifying pulmonary tuberculosis[7]. In ophthalmic research, application of AI technology have led to excellent diagnostic accuracy for several ocular conditions such as diabetic retinopathy (DR), age related macular degeneration (AMD), and sickle cell retinopathy (SCR).[2, 4, 8-10]

In the current clinical setting, mass screening programs for common ocular conditions such as DR or SCR are heavily dependent upon experienced physicians to examine and evaluate retinal photographs. This process is time consuming and expensive, making it difficult to scale up to incorporate the millions of individuals, who harbor systematic diseases which are prone to affect the retina. Patients with early onset of retinopathies such as DR or SCR are initially asymptomatic yet require monitoring to ensure prompt medical interventions to prevent vision losses. However, it is not feasible to screen 65 million people in the USA over the age of 50 years,[1] to identify for individuals with signs of early retinopathy (AMD, DR or other disease). An AI-based diagnostic tool with capability for multiple-disease differentiation would have tremendous potential to advance mass-level screening of eye diseases.[11]

To date, most of the reported studies of AI diagnostic systems in literature are based on color fundus photography.[12-15] Fundus photography is one of the most common clinical imaging modalities and has been widely used in evaluating retinal abnormalities. Supervised and unsupervised machine learning based diagnostic systems using fundus images have been developed by researchers for staging of individual retinopathies as well as to identify multiple ocular diseases.[8, 16-19] However, these demonstrated AI-based diagnostic tools generally face two major challenges. Firstly, fundus images provide limited resolution and retinal vascular information, limiting its capability to quantify subtle micro-vascular distortions near the foveal area and in different retinal layers. Thus, diagnostic systems using supervised machine learning algorithms suffer from low-performing quantitative feature analysis and concurrently low diagnostic accuracy. Secondly, systems using unsupervised or deep machine learning require a large and well documented database (ranging from 100,000 to millions) for training and optimizing convolutional neural networks. Even if an AI system is successfully trained, the intrinsic variance among different database from multiple imaging centers makes it extremely difficult to provide robust accuracy metrics. Additionally, in case of new retinal imaging modalities such as OCTA, it is quite challenging to accumulate large, multi-center database for efficient clinical deployment of AI-based diagnostic tools.

As a potential solution to overcome these challenges, we propose a supervised machine learning based approach to train and evaluate a support vector machine (SVM) classifier model with quantitative optical coherence tomography (OCT) angiography (OCTA) features for multi-task AI classification of retinopathies. By providing excellent capability for depth-resolved visualization of retinal vascular plexuses, quantitative OCTA holds genuine promise for AI screening of retinopathies. Although the comparatively smaller data size of OCTA presently limits deep-learning based strategies, the sensitivity of OCTA features to detect onset and progression or retinopathies make it readily useful for supervised AI based screening. Recent studies have established several quantitative OCTA features correlated with subtle pathological and microvascular distortions in the retina. OCTA features such as blood vessel tortuosity (BVT), blood vascular caliber (BVC), vessel perimeter index (VPI), blood vessel density (BVD), foveal avascular zone (FAZ) area (FAZ-A), and FAZ contour irregularity (FAZ-CI) have also been validated for objective classification and staging of DR[5, 20] and SCR,[21] individually. Our recent studies demonstrated that DR and SCR show different effects on OCTA features, and thus quantitative OCTA analysis promises the potential of multiple-task classification to differentiate retinopathies and stages. In this study, we propose to test the feasibility of using these quantitative OCTA features for machine leaning based multi-task AI screening of different retinopathies. For easy comparison with our recent studies, DR and SCR were selected as the two diseases for technical validation of the proposed AI screening methodology. The AI system containing a SVM classifier model utilizes a hierarchical backward elimination technique to identify optimal-feature-combination for the best diagnostic accuracy and most efficient classification performance. The AI-based screening tool performs multi-layer hierarchical tasks to perform 1) normal vs. disease classification, 2) inter-disease classification (DR vs.

SCR), and 3) staging of DR (mild, moderate and severe non-proliferative DR (NPDR)) and SCR (mild and severe). The performance of the AI system has been quantitatively validated with manually labeled ground truth, using sensitivity, specificity and accuracy metrics along with graphical metrics, i.e., receiver operation characteristics (ROC) curve.

## Results

The OCTA image database in this study included 115 images from 60 DR patients (20 mild, 20 medium and 20 severe NPDR), 90 images from 48 SCR patients (30 stage II mild and 18 stage III severe SCR), and 40 images from 20 control patients (representative images shown in supplementary Fig S1). Patient demographic data is shown in table 1. There were no statistical significances in age and sex distribution of between control, DR and SCR groups. (ANOVA, P = 0.14; chi-square test, P = 0.11 and P = 0.32, respectively). For DR, no significance in hypertension or insulin dependency between stages of disease groups was observed.

Table 1. Demographics of control, DR and SCR subjects

|  | Control | DR | | | SCR | |
| --- | --- | --- | --- | --- | --- | --- |
|  |  | Mild NPDR | Moderate NPDR | Severe NPDR | Mild SCR | Severe SCR |
| Number of subjects | 20 | 20 | 20 | 20 | 30 | 18 |
| Sex (male) | 12 | 11 | 12 | 11 | 17 | 11 |
| Age (mean ± SD) | 42 ± 9.8 | 50.1 ± 12.61 | 50.8 ± 8.39 | 57.84 ± 10.37 | 51 ± 11.52 | 59.73 ± 8.26 |
| Age range | 25-71 | 24-74 | 32-68 | 41-73 | 28-71 | 46-75 |
| Duration of disease | - | 19.64±13.27 | 16.13±10.58 | 23.40 ± 11.95 | 13.25± 8.78 | 18.43±10.7 |
| Diabetes type | - | Type II | Type II | Type II | - | - |
| Insulin dependent(Y/N) | - | 7/13 | 12/8 | 15/5 | - | - |
| HbA1C, % | - | 6.5 ± 0.6 | 7.3 ± 0.9 | 7.8 ± 1.3 | - | - |
| HTN prevalence, % | 10 | 45 | 80 | 80 | - | - |

[a] DR, diabetic retinopathy, SD, standard deviation, HbA1C, Glycated hemoglobin, HTN, hypertension

**Optimal feature selection using backward elimination**

We employ a logistic regression based model with backward elimination to select optimal combination of features for the multi-task classification. A summary of the quantitative univariate analysis of the OCTA features is shown in supplementary Table S1-S3 for comparing control vs. DR vs. SCR, NPDR stages and SCR stages respectively. In general, BVT, BVC and FAZ parameters increased with disease onset and progression whereas BVD and VPI decreased. The comparison of the diagnostic accuracy for each feature in the backward elimination process is shown in table 2. Figure 1 provides further support to the results shown in table 2, showing relative changes of OCTA features in different groups. Each panel corresponds to four classification tasks respectively. The backward elimination initially started with all OCTA features and eliminated features one by one based on the prediction accuracy of the fitted regression model. The feature selection method identified an optimal feature combination for each classification task, i.e. perifoveal $BVD_{SC3}$ (SCP, circular area: > 4mm), $FAZ\text{-}A_S$ (SCP) and $FAZ\text{-}CI_D$ (DCP) for control vs. disease classification; $BVT_S$ (SCP), $BVD_{SC3}$, $FAZ\text{-}A_S$, and $FAZ\text{-}CI_D$ for DR vs. SCR classification; $BVD_{SC3}$ and $FAZ\text{-}A_S$ for NPDR staging; and $BVT_S$, $BVD_{SC3}$, and $FAZ\text{-}CI_S$ (SCP) for SCR staging. From table 2, we can observe that the individual accuracy of the optimal features in each classification tasks were highest compared to the other features and the model fitted with the combination of these optimal features provided the best diagnostic accuracy. Also, from figure 1, we can see that the relative changes in each cohort can be only observed in the chosen optimal OCTA features.

**Multi-task classification**

The SVM classifier performs the classification tasks in a hierarchical manner. To evaluate the diagnostic performance in each step or task, we measured the sensitivity and specificity task. For each task, the ROC curves were also drawn (Fig. 2) and AUCs were calculated. At the first step, the SVM identifies diseased patients from control subjects with 97.84% sensitivity and 96.88% specificity (AUC 0.98). After identifying the diseased patients, the classifier sorts them to two groups: DR and SCR with 95.01% sensitivity and 92.25% specificity (AUC.94). After sorting to corresponding retinopathies, the SVM conducted the condition staging classification: 92.18% sensitivity and 86.43% specificity for NPDR staging (mild vs. moderate vs. severe; AUC 0.96), and 93.19% sensitivity and 91.60% specificity for SCR staging (mild vs. severe; AUC 0.97). The sensitivity, specificity and AUC metrics are calculated for SVM model trained with optimal feature combination. Supplementary table S4 shows the performance metrics in further details.

Table 2. Diagnostic accuracy measured during hierarchical backward elimination.

| Parameters | Diagnostic accuracy (%) | | | |
|---|---|---|---|---|
| | Control vs. Disease | DR vs. SCR | NPDR Staging | SCR Staging |
| $BVT_S$ | 81.75 | 81.64 | 71.26 | 89.15 |
| $BVC_S$ | 79.88 | 75.59 | 78.51 | 71.92 |
| $VPI_S$ | 76.49 | 76.83 | 78.39 | 65.46 |
| $BVD_{SC1}$ | 72.11 | 53.14 | 62.02 | 55.19 |
| $BVD_{SC2}$ | 80.02 | 77.98 | 75.83 | 74.98 |
| $BVD_{SC3}$ | 89.01 | 83.49 | 82.67 | 83.67 |
| $BVD_{DC1}$ | 69.35 | 52.17 | 64.30 | 58.02 |
| $BVD_{DC2}$ | 78.53 | 75.83 | 78.54 | 76.20 |
| $BVD_{DC3}$ | 80.69 | 70.28 | 77.13 | 65.59 |
| $FAZ\text{-}A_S$ | 91.67 | 83.66 | 85.02 | 78.84 |
| $FAZ\text{-}A_D$ | 88.48 | 80.09 | 80.46 | 76.11 |
| $FAZ\text{-}CI_S$ | 88.74 | 81.57 | 79.34 | 80.95 |
| $FAZ\text{-}CI_D$ | 89.05 | 82.65 | 78.95 | 75.69 |
| **Optimal feature combination** | **97.45** | **94.32** | **89.60** | **93.11** |

[a] Superscript S and D denote SCP and DCP respectively. In case of BVD, C1-C3 denote circular area 1,2 and 3 respectively as shown in figure 5.

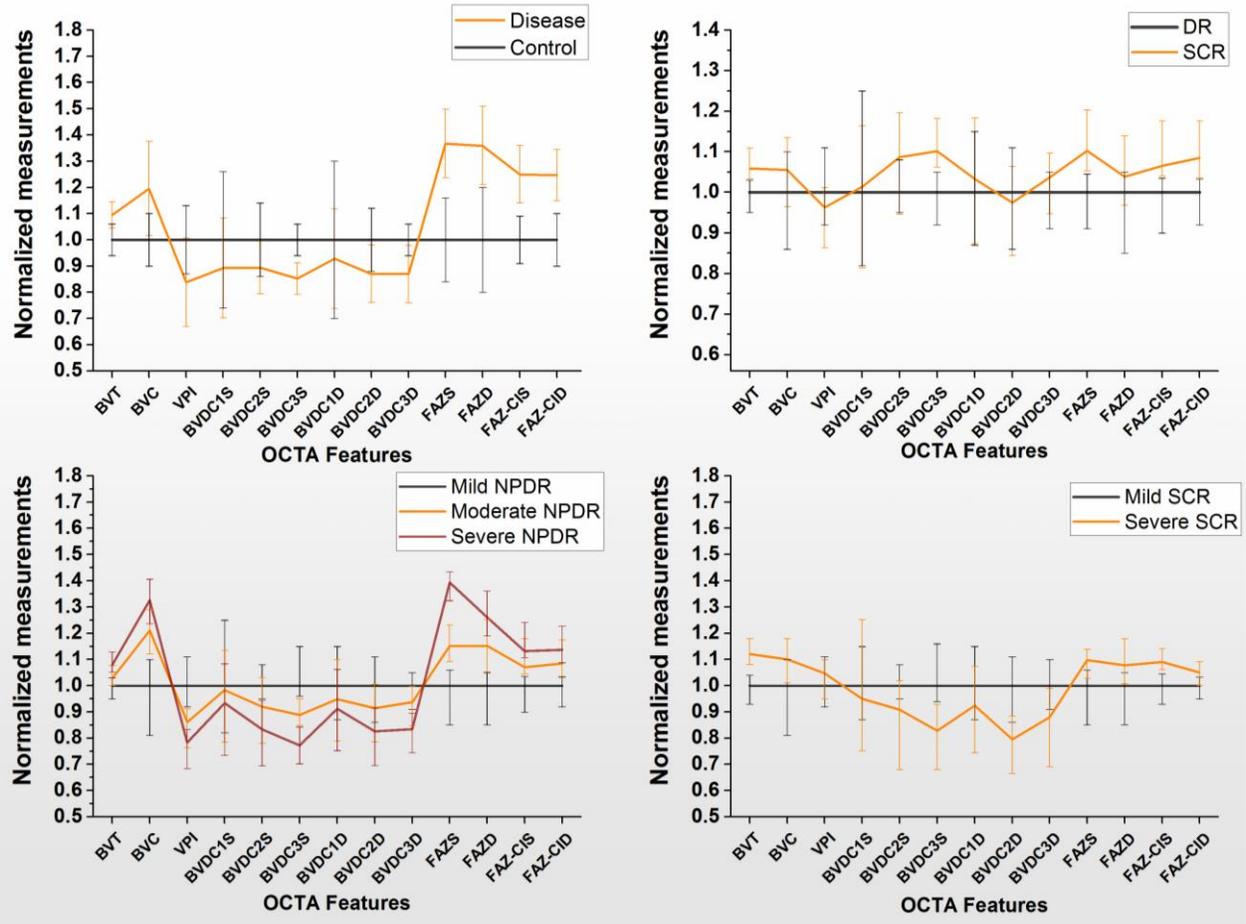

Figure 1: Normalized feature trends for different cohorts. (A) Change in Disease group (DR and SCR) compared to control. (B) Change in SCR compared to DR. (C) Change in moderate and severe NPDR compared to mild NPDR. (D) Change in severe SCR compared to mild SCR. Error bars represent standard deviation.

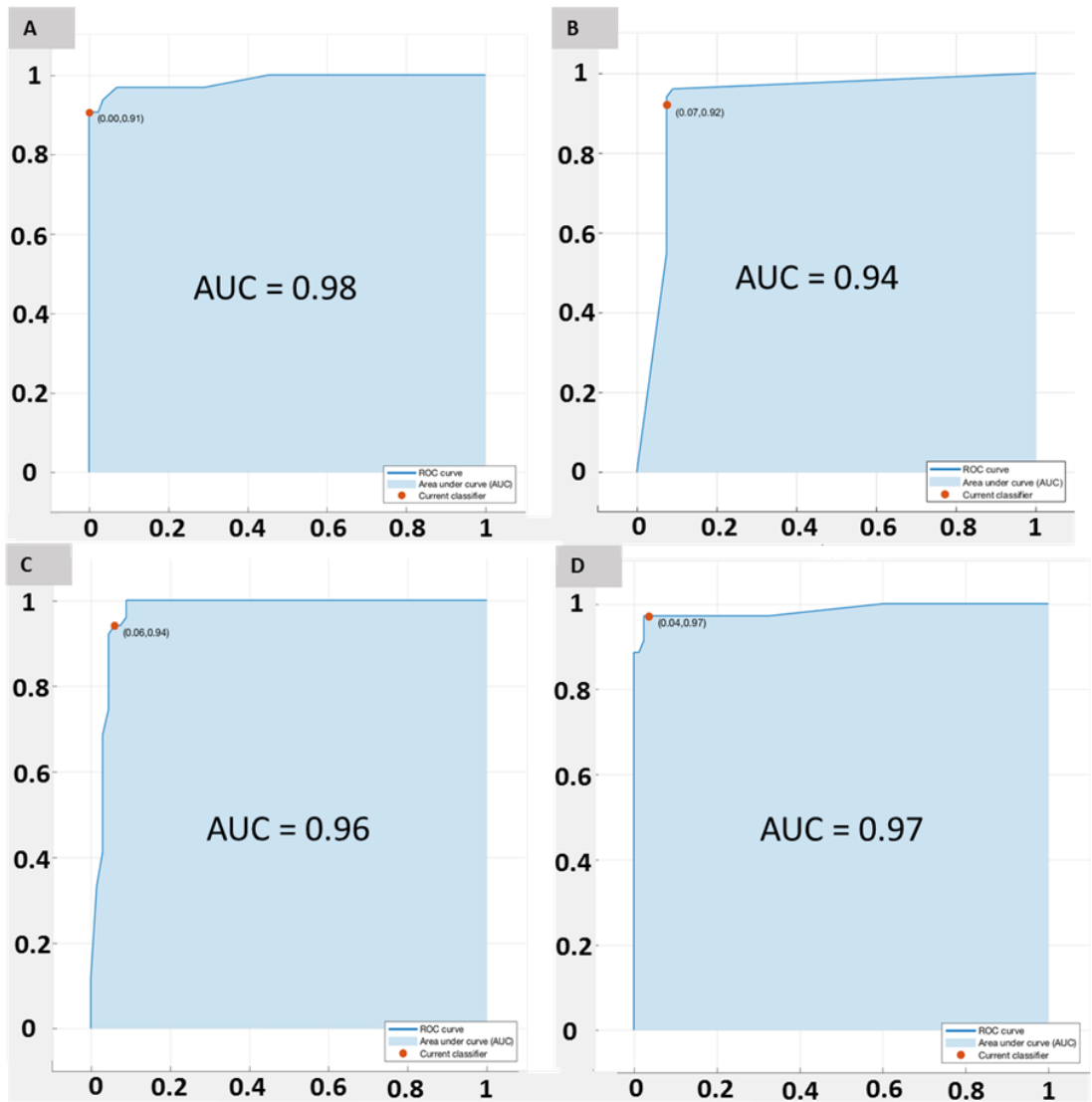

Figure 2. ROC curves illustrating classification performances of the prediction model using optimal combination of features. (A) Control vs disease classification. (B) DR vs. SCR classification. (C) NPDR staging. (D) SCR staging.

## Discussion

We herein demonstrate the feasibility of a supervised machine learning based AI screening tool for multiple retinopathies using quantitative OCTA technology. In a hierarchical manner, this diagnostic tool can perform multiple tasks to classify i) control vs. disease, ii) DR vs. SCR, iii) different stages of NPDR and SCR, using quantitative features extracted from OCTA images. OCTA images can provide visualization of subtle microvascular structures in intraretinal layers which permits a comprehensive quantitative analysis of pathological changes due to systematic retinal diseases such as DR and SCR.

Morphological distortions such as impaired capillary perfusion, vessel tortuosity and overall changes in foveal size and complexity etc. were quantitatively measured and compared for identifying onset and progression of DR or SCR in diabetes and SCD patients respectively. The SVM classifier model demonstrated a robust diagnostic performance in all classification tasks. The classification model also utilized a back-elimination strategy for choosing an optimal combination of OCTA features for getting the best diagnostic performance with highest efficiency. Proper implementation of this AI-based tool in primary care centers would facilitate a quick and efficient way of screening and diagnosis of vision impairment due to systematic diseases.

For any screening and diagnostic prediction system, sensitivity is a patient safety criterion.[22] The AI-based tool's major role is to identify patients prone to vision impairment due to retinopathies. In the control vs disease classification task, 94.84% sensitivity of our system represents the capability to identify individual eyes with retinopathies (DR and SCR) from a general pool of control, DR and SCR eyes. Furthermore, the system can identify patients with DR or SCR with 95.01% sensitivity. This is crucial for screening purposes, as those patients should be referred to eye care specialists. Similarly, specificity is also an important factor because it will represent the capability of detecting subjects that don't require referral to an eye care specialist. When the data pool equals millions of patients, this discriminatory capability is crucial for efficient clinical effectiveness in mass-screening. Our system demonstrates 96.88% specificity which means the control subjects would rarely be erroneously referred for treatment of retinopathies; additionally, 92.25% specificity in DR vs. SCR classification means the patients with DR or SCR would not be referred with an incorrect diagnosis. This is relevant since certain advanced stages of a disease tend to progress faster than others and hence require more expedient evaluation and management upon referral. In mass-screening applications, the AI classification tool will be useful to identify proper referral for patients with systematic diseases (i.e. diabetes or SCD) and avoid unnecessary referral for patients who don't need specialized care at that time point.

Our study demonstrated that an optimal combination of OCTA features can achieve maximum diagnostic accuracy for all classification tasks. As supported by results from table 2 and figure 1, we can observe that, in all performance metrics, the classification model trained with optimal feature combination demonstrated better diagnostic proficiency compared to the model trained with individual features or combination of all features. The OCTA features analyzed in this study represent vascular and foveal distortions in retina due to retinopathy from both superficial and deep layers as well as localized circular regions in the retina (BVD). Out of all these OCTA features, the feature selection strategy identified the most sensitive features for each classification tasks to significantly distinguish different cohorts.] The high diagnostic accuracy of SVM classifier trained with optimal feature combination highlights the importance of the most relevant feature selection in automated classification. Few features that showed significance in the univariate analysis (Supplementary tables) were not selected in the final set of optimal features. This suggests a contrast between clinical applicability and overall difference of OCTA features among different patient groups. Ashraf et. al.[20] observed a similar phenomenon when using feature selection for automated staging of DR eyes. In all the classification tasks, the most sensitive features also had low correlation amongst themselves. Figure 3 illustrates a scatter plot showing correlation analysis for DR vs. SCR classification. We can observe that only FAZ parameters had positive correlation with each other; BVT and BVD both were not significantly correlated with FAZ parameters (Spearman's rank test, $P>0.05$), suggesting that all the features provided different pathological aspects of the diseased retina. Therefore, the four optimal features were objective for identifying DR or SCR associated distortions and their combination yielded strong classification performance.

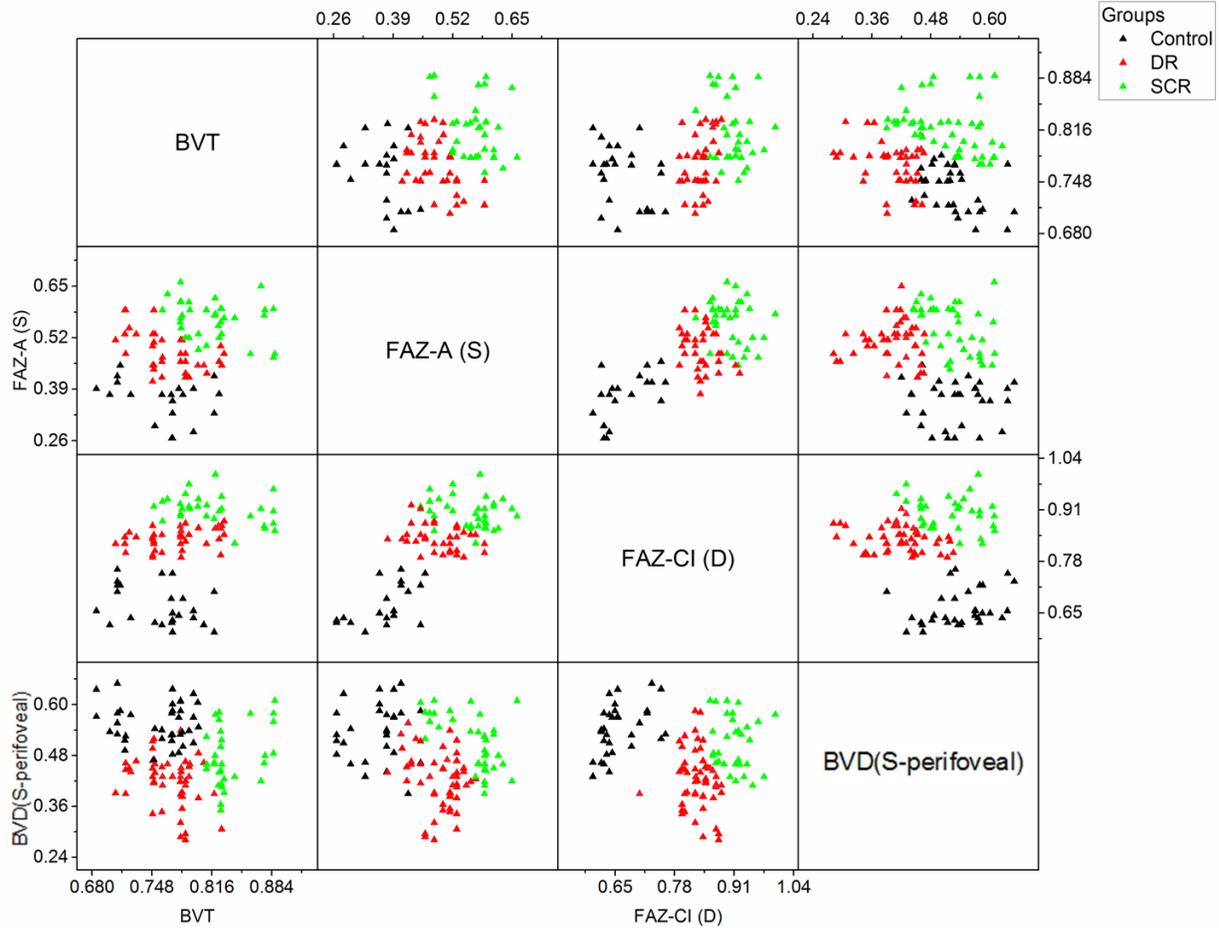

Figure 3. Correlation analysis among four most sensitive features. The scatter plot also shows the distribution of control, DR and SCT patient data for different feature combination.

The optimal OCTA features selected by the AI classification tool have been previously shown in the literature to be useful in quantitative analysis studies..[23-33] Both BVD and FAZ parameters (FAZ-A and FAZ-CI) have been shown to be significant in identifying DR stages.[20, 31, 33, 34] BVT is also an established predictor for SCR progression. In two separate studies, we previously demonstrated an SVM classifier for automated staging of DR groups (mild, moderate, severe)[5] and SCR groups (stage II and III).[21] In our DR study, the most sensitive OCTA feature was observed to be BVD while for SCR, it was BVT and FAZ. These sensitive OCTA features are also selected to be included in the optimal feature set by the backward elimination technique in our current study for different classification tasks. Our current study, therefore, supports our previous findings and also demonstrates the clinical importance of identifying most sensitive features for different retinopathies. Furthermore, the optimal features included measurements from both SCP and DCP. Previous OCTA studies[20] including our recent studies[5, 35] have suggested that the onset and progression of DR or SCR in diabetes or SCD patients affect both the retinal layers. By choosing optimized features from SCP and DCP, the AI-based model ensured representation of layer specific distortions due to retinopathies.

For practical implementation of any AI-based tool in mass-screening at a clinical setting, a major challenge is the computation time required for overall feature extraction, optimization and diagnostic prediction. Our AI-based screening tool required only 4-6 seconds to extract features from each OCTA image. From the training data, the optimized features are chosen using back elimination which takes approximate 40-50 seconds (done only one time) depending on the size of the dataset. After the training of the SVM classifier is completed, it takes 8-10 seconds for classifying the testing database used in this study. If new data is included for diagnosis prediction, it takes only 1-2 seconds per OCTA image to use the trained model to classify control, DR or SCR eyes. However, at this point the AI-based tool is implemented in MATLAB (Mathworks, Natick, MA, USA), a separate software not integrated in the OCTA imaging device (Angiovue from Optovue, Fremont, CA, in our case). Once the technology is integrated into the interface of the OCTA device, the users can view real-time prediction as soon as the OCTA image is captured in retina clinics. The diagnostic accuracy can be enhanced even further if the patient history or clinical information is integrated into the screening tool.

Limitations of this study include relatively modest sample size for each of cohort and single imaging center. In future studies, we plan to include multiple imaging centers and a much larger OCTA database to test the robustness of our AI screening tool for practical implementation in retina clinics. Furthermore, we relied on the segmentation provided by the clinical device to identify the images from SCP and DCP. Thus, there is a possibility of segmentation error. The potential motion, projection artifacts in OCTA and error in reconstruction of OCTAs from SD-OCT volume data were few other limitations. However, we attempted to minimize the effect of these errors and artifacts in our study by excluding the images with severe artifacts, segmentation errors and patients with macular edema.

In conclusion, we present a supervised machine learning based multi-task AI classification tool that uses an optimal combination of quantitative OCTA features for objective classification of control, DR and SCR eyes with excellent diagnostic accuracy. Using the feature selection strategy, the classifier selected $BVD_{SC3}$, $FAZ\text{-}A_S$ and $FAZ\text{-}CI_D$ for control vs. disease classification; $BVT_S$, $BVD_{SC3}$, $FAZ\text{-}A_S$, and $FAZ\text{-}CI_D$ for DR vs. SCR classification; $BVD_{SC3}$ and $FAZ\text{-}A_S$ for staging of NPDR severity; and $BVT_S$, $BVD_{SC3}$, and $FAZ\text{-}CI_S$ for staging of SCR severity . The optimal-feature-combination directly correlates to the most significant morphological changes in retina for each classification tasks and provides the most effective classification performance with least computational complexity. Our diagnostic tool performs well with cross-validate data. However, further validation studies using larger cohorts of OCTA data from different centers and devices will facilitate future clinical implementation of a mass-level AI-based screening tool.

## Methods

Figure 4 illustrates the step by step methodology for the machine learning based multi-task AI classification. Each classification task involves primarily three steps. The first step is OCTA image data acquisition and feature extraction (DA&FE). The second step is optimal feature identification (OFI) using a hierarchical backward elimination technique for the specific classification task. The third step is to validate multiple-task classification (MTC) using the identified optimal-feature-combinations.

**Data Acquisition and feature extraction**

*OCTA data acquisition:* This cross-sectional study was approved by the Institutional Review Board (IRB) of the University of Illinois at Chicago (UIC) and complied with the ethical standards stated in the Declaration of Helsinki. Both the DR and SCR patients were recruited from UIC Retinal Clinic. All patients underwent complete anterior and dilated posterior segment examination (JIL, RVPC). For DR, a retrospective study of consecutive type II diabetes patients was conducted who underwent OCT/OCTA imaging. The patients are representative of a university population of diabetic patients who require imaging for management of diabetic macular edema and DR. Two board-certified retina specialists classified the patients based on the severity of DR (mild, moderate, severe NPDR) according to the Early Treatment Diabetic Retinopathy Study (ETDRS) staging system. In case of SCR, disease stages were graded according to the Goldberg classification (stage I-V, from mild to severe). Only stage II (mild) and III (severe) SCR data were included in this study as stage I OCTA data were limited in number while stage IV OCTA images were unreliable due to distortions caused by hemorrhages and vessel proliferation. For simplification in the classification process, we define the stage II and III as mild and severe stage SCR, respectively. The control OCTA data were obtained from healthy volunteers who gave informed consent for OCT/OCTA imaging. Both eyes (OD and OS) were examined and imaged. We did not include eyes with other ocular disease or any pathological features in their retina such as epiretinal membranes and macular edema. Additional exclusion criteria included eyes with prior history of vitreoretinal surgery, intravitreal injections or significant (greater than a typical blot hemorrhage) macular hemorrhages.

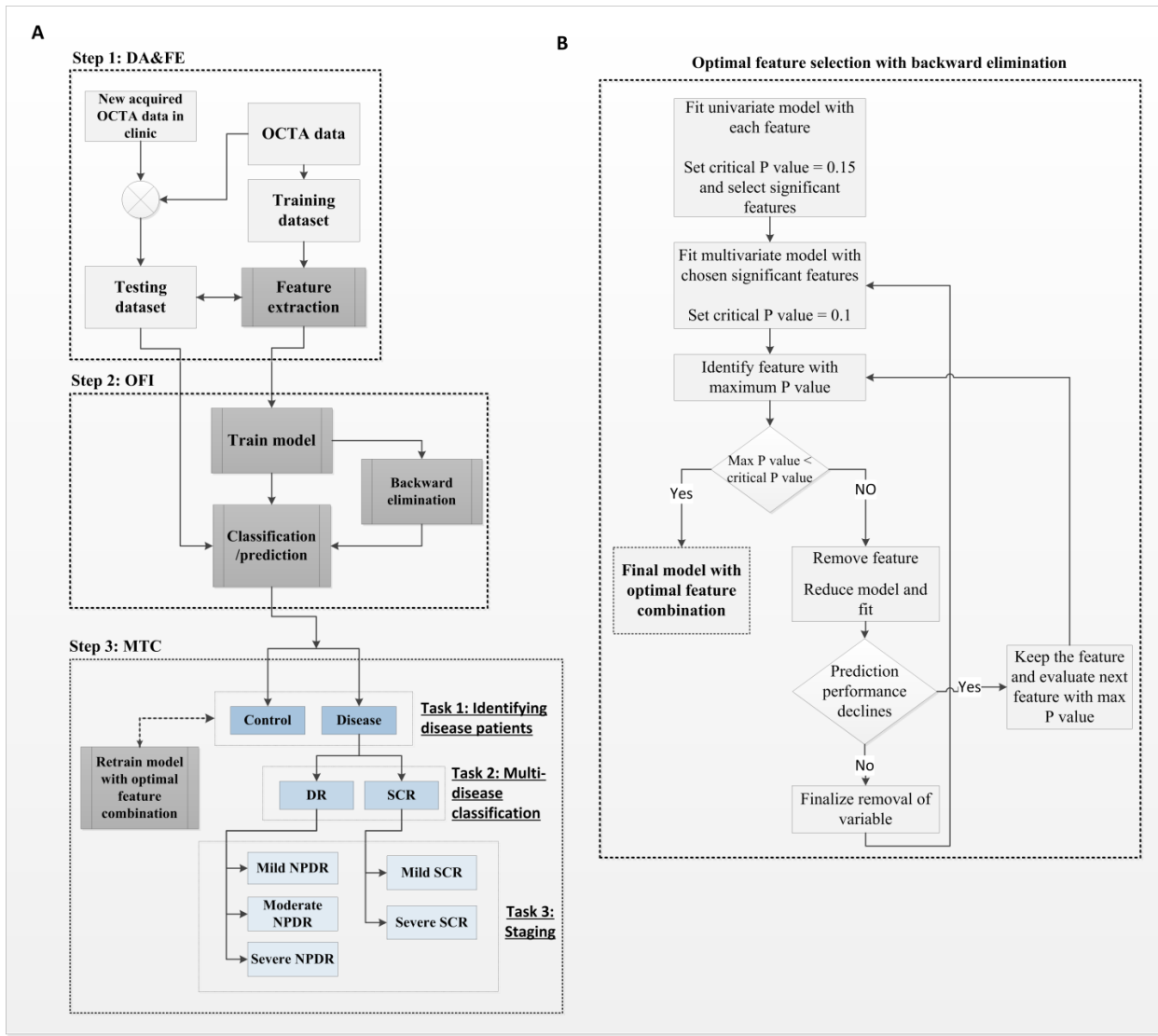

Figure 4: (A) Step by step methodology of AIbased classification. (B) Optimal feature selection with hierarchical backward elimination technique. DA&FE: Data acquisition and feature extraction; OFI: Optimal feature identification; MTC: Multiple-task classification.

SD-OCT and OCTA image data were acquired using an Angiovue SD-OCT device (Optovue, Fremont, CA, USA), consisting of a 70,000 Hz A-scan rate, and axial and lateral resolutions of ~5 μm and ~15 μm, respectively. All OCTA images used in this study were 6 mm×6 mm scans; OCTA images were acquired from both superficial and deep capillary plexuses (SCP and DCP). All the images were qualitatively examined, OCTA images with severe motion or shadow artifacts were also excluded. The OCTA images were exported from imaging device and custom-developed MATLAB procedures were used for image processing, feature extraction and classification as described below.

***OCTA pre-processing and feature extraction:*** All the OCTA images used in this study had a field of view (FOV) of 6 mm × 6 mm (304×304 pixels). The OCTA images were normalized to a standard window level based on the maximum and minimum intensity values to account for light and contrast image variation. Bias field correction and contrast adjustment of the OCTA images improved the overall reliability of the extracted features and concurrently the performance of classifier model to identify OCTAs from different cohorts.

Six different quantitative OCTA features were extracted from each OCTA image (Fig. 5) for the AI classification. The vascular features were BVT, BVC, VPI, and BVD, while the foveal features were FAZ-A and FAZ-CI. Before measuring the vascular features, the vessel map and skeleton map were extracted from the OCTA image (Fig 5B and 5C). For the vessel map, we have used a Hessian based multi-scale Frangi filter[36] to enhance vascular flow information. This method utilizes the Eigen vectors of the Hessian matrices and calculates the likeliness of an OCTA region to be vascular structures. Adaptive thresholding along with morphological functions were furthers used for cleaning the vessel map and removing noises. From the vessel map, a skeleton map is generated using morphological shrinking functions. The extracted vessel and skeleton maps from OCTA images had an average area of 47.34% and 25.81% respectively.

A brief description of the feature measurement procedure is as follows:

**BVT**: The BVT was measured in the SCP. For BVT measurement, the BVT of each vessel branch is measured from the skeleton map and average BVT is measured as,

$$\text{BVT} = \frac{1}{n}\sum_{i=1}^{n}\left(\frac{\text{Geodesic distance of a vessel branch i}}{\text{Euclidean distance of a vessel branch i}}\right) \quad (1)$$

$$\text{Euclidean distance} = \sqrt{(x_1 - x_2)^2 + (y_1 - y_2)^2} \quad (2)$$

$$\text{Geodesic distance} = \int_{t_0}^{t_1}\sqrt{\left(\frac{dx(t)}{dt}\right)^2 + \left(\frac{dy(t)}{dt}\right)^2}\, dt \quad (3)$$

where [$x_i, y_i$] are the two endpoints of a vessel branch.

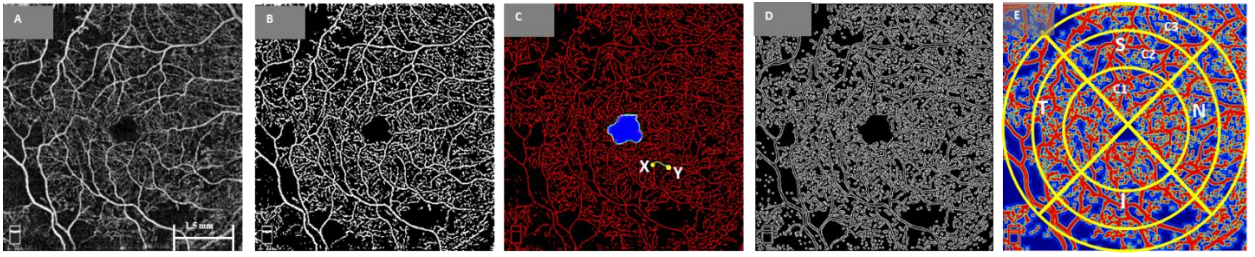

Figure 5. Representative OCTA images for illustrating the feature extraction. (A) Original OCTA image from a severe NPDR patient. (B) Segmented blood vessel map including large blood vessels and small capillaries. Hessian based Frangi vesselness filter and FD classification provide a robust and accurate blood vessel map. (C) Blood vessel skeleton map (red) with segmented fovea (marked blue region) and FAZ contour (green boundary marked around fovea). A random vessel branch is highlighted in green with X and Y endpoints identified in yellow dot. (D) Vessel perimeter map. (E) Fractal contour maps for blood vessel density measurement. Scale bar shown in A applies to all the images.

**BVC**: BVC was measured from the SCP as the ratio of vascular area (calculated from vessel map) and vascular length (calculated from skeleton map),

$$\text{BVC} = \frac{\text{Vascular area}}{\text{Vascular length}} \quad (4)$$

**VPI**: VPI was measured from the perimeter map (Fig. 5D) in SCP as the ratio of vessel perimeter area and total image area,

$$\text{VPI} = \frac{\text{Perimeter area}}{\text{Total image area}} \quad (5)$$

**BVD**: BVD was measured in both SCP and DCP using fractal dimension (FD) technique. The details and rationale about FD calculation is previously described.[35] Each pixel is assigned an FD value from 0 to 1 where 0 corresponds to avascular region and 1 corresponds to large vessel pixels. The FD of 0.7 to 1 corresponds to vessel pixels and average BVD is measured as the vascular area to total image area.

$$\text{BVD} = \frac{\text{Vascular area}}{\text{Total image area}} \quad (6)$$

BVD the measurements were taken in three localized regions in the retina, three circular regions of diameter 2 mm, 4 mm and 6 mm (C1, C2, and C3) around the fovea (as shown in Fig.5E). The segmented FAZ area was excluded when measuring BVD for improved diagnostic accuracy.

**FAZ-A**: The FAZ-A was measure in both SCP and DCP. The fovea was demarcated automatically (blue area in Fig. 5C)[35] and FAZ-A was measured as,

$$\text{FAZ} - \text{A } (\mu m^2) = \text{Number of pixels is Fovea} \times \text{Area of single pixel} \qquad (6)$$

The automatically segmented FAZ area was compared to manually traced FAZ labelling and had 98.26% similarity with manually segmented ground truths.

**FAZ-CI**: FAZ-CI was measured in both SCP and DCP. From the demarcated fovea, FAZ contour was segmented automatically[35] (green demarcated contour in Fig.5C). From the segmented contour the FAZ-CI was measured as,

$$\text{FAZ} - \text{CI} = \frac{\text{Perimeter of the FAZ contour}}{\text{Perimeter of a circle with equivalant area to the FAZ}} \qquad (5)$$

**Optimal feature identification**

*Statistics and classification model:* Statistical analyses were conducted using MATLAB (Mathworks, Natick, MA, USA) and OriginPro (OriginLab Corporation, MA, USA). All the OCTA features were tested for normality using a Shapiro-Wilk test. For normally distributed variables, one versus one comparisons were conducted using Student's t-test and one way, multi-label analysis of variance (ANOVA) was used to compare differences among multiple groups. If the features were not normally distributed, we used independent sample t-test (Mann-Whitney) for one versus one comparisons and non-parametric Kruskal-Wallis test for comparing multiple groups. Chi-square test was used to compare the sex and hypertension distribution among different groups. For age distribution, we used ANOVA. Spearman's correlation coefficients ($r_s$) were measured to analyze the relationship among the OCTA features and their correlation with DR or SCR severity. Statistical significance for univariate analysis and correlation test was defined with $P < 0.05$; however, the P values were Bonferroni-corrected for multiple simultaneous group comparisons. For the classification model that would be trained with OCTA features and perform the diagnosis prediction, we chose a support vector machine (SVM) classifier. In case of logistic regression based backward elimination (Fig. 4B), the initial critical value of P was 0.15 for the univariate model while it was 0.1 for multi-variate model. In this case, a P value of 0.05 or less is too conservative and there might be a possibility of losing valuable information from multivariate regression analysis of different features.

*Optimal feature selection with backward elimination:* We implemented feature optimization to choose a subset of OCTA features that delivers the best diagnostic prediction for each classification tasks, i.e. 1) identifying disease patients from control, 2) inter-disease (DR vs. SCR) classification and 3) Staging of DR (mild, moderate, and severe NPDR) and SCR (mild and severe) respectively. Taking inspiration from Occam's Razor, we aim to choose the smallest classification model that fits the data. For choosing this optimal feature combination for each classification task, we used a stepwise backward elimination technique. The flowchart of necessary steps taken in backward elimination of features is illustrated in the Fig.4B. Backward elimination starts with all of the predictors in the model. The variable that is least significant--that is, the one with the largest P value with worst prediction performance in a regression analysis is removed and the model is refitted. Each subsequent step removes the least significant variable in the model until all remaining variables have individual P values smaller than critical P value (set at 0.05). After the SVM is trained with the optimal feature combination, we tested the classification model with a testing data set. This feature selection process using backward elimination is repeated for each of the steps and the SVM model is trained with corresponding optimal feature combination at each step for a specific classification task. For control vs. disease and DR. vs. SCR classification, the SVM performs a binary (one vs one) classification while for staging disease conditions (mild vs. moderate vs. severe NPDR and mild vs. severe SCR) the SVM performs a multi-class classification. The prediction is performed on the testing database with 5-fold cross validation to control any overfitting. Once the SVM is trained with optimal feature combination, any new data can be directly inputted into the classifier to generate task-specific predictions.

*Performance metrics:* The performance of the prediction model was evaluated with sensitivity, specificity, and accuracy metrics. ROC (Receiver Operation Characteristics) curves were also generated along with area under the ROC curve (AUC). ROC curve plots the true positive rate (i.e., sensitivity) as a function of false positive rate (i.e., 1-specificity) at different tradeoff points. AUC is measured to quantify how well the classifier is able to identify the different classes. The closer the curve to the left up corner, the more accuracy the prediction is. AUC equals 1 or 100% represents a perfect prediction, and 0.5 or 50% represents a bad prediction.

**Data and materials availability**

The datasets generated during the current study that were used to calculate the primary outcome parameters are available upon reasonable request from the corresponding author, X.Y.

**Code availability**

The AI system described in this study is an SVM model, which is already established classifier model. The feature extraction codes have been previously described in our published articles. However, they can be made available upon reasonable request form X.Y.

**Acknowledgement**

The authors thank Mr. Mark Janowicz and Ms. Andrea Degillio (Eye and Ear Infirmary, University of Illinois at Chicago) for technical support of data acquisition.

**Financial support:** This research was supported in part by NIH grants R01 EY030101, R01 EY024628, P30 EY001792; by unrestricted grant from Research to Prevent Blindness; by Richard and Loan Hill endowment; by Marion H. Schenk Chair endowment.

**Author contributions**

M. A. contributed to algorithm development, data acquisition, and manuscript preparation; D.L. contributed to data acquisition and manuscript preparation; J.I.L and R.V.P.C. contributed to study design and manuscript preparation; X.Y. supervised this project, and contributed to study design, data analysis and manuscript preparation.

# **Supplementary materials**

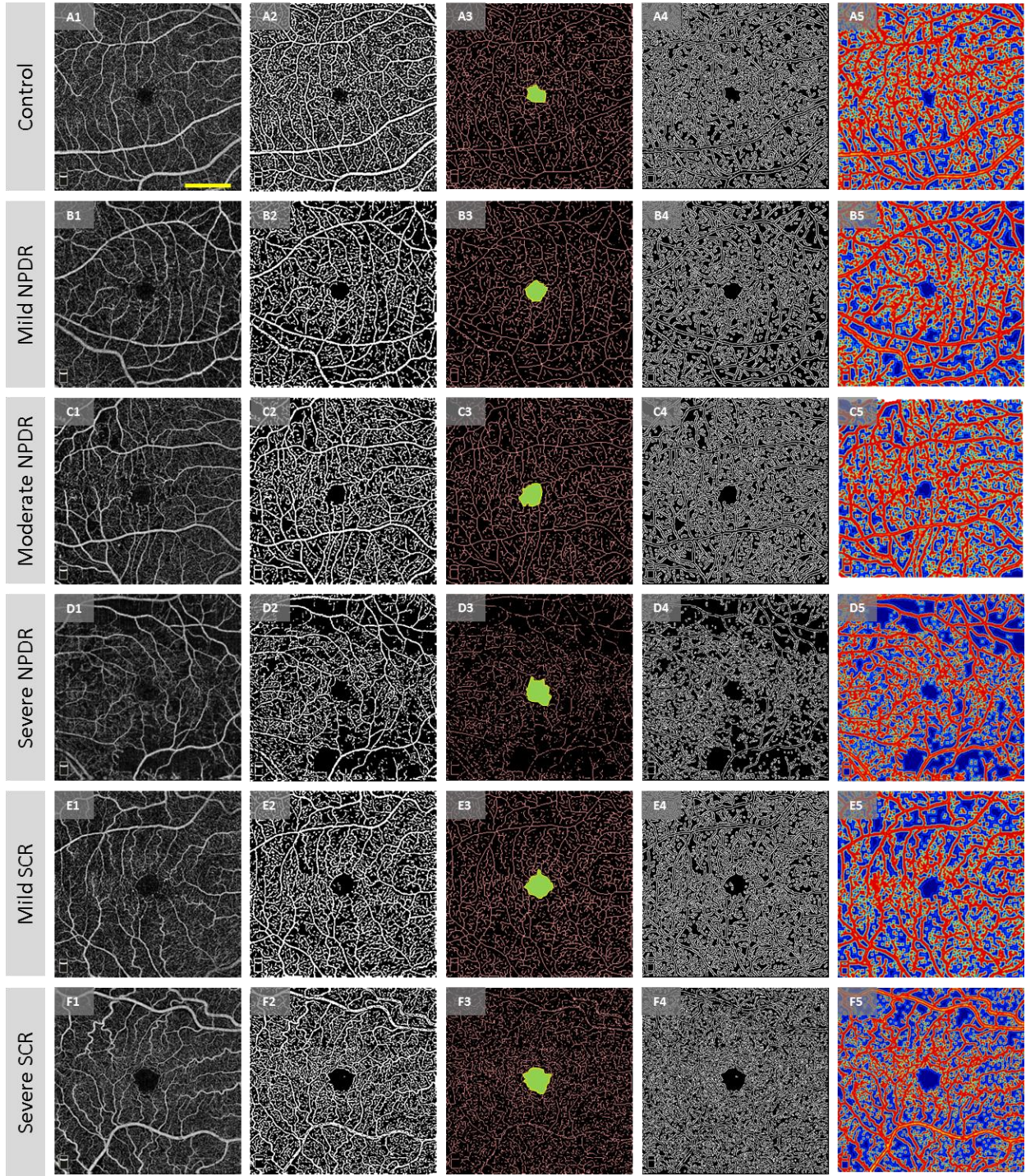

Figure S1. Representative OCTA images for illustrating the feature extraction. (A1-A5) Control subject, (B1-B5) Mild NPDR subject, (C1-C5) Moderate NPDR subject, (D1-D5) Severe NPDR subject, (E1-E5) Mild SCR (stage II) subject, (F1-F5) Severe SCR subject. (Column 1) OCTA image. (Column 2) Segmented blood vessel map including large blood vessels and small capillaries. Hessian based Frangi vesselness filter and FD classification provide a robust and accurate blood vessel map. (Column 3) Skeletonized blood vessel map (red) with segmented FAZ (marked green region) and FAZ contour (yellow boundary marked around FAZ). (Column 4) Vessel perimeter map. (Column 5) Contour maps created with normalized values of local fractal dimension. Scale bar shown in A1 corresponds to 1.5 mm and applies to all the images.

# Univariate analysis

Table S1. Univariate analysis of individual OCTA features for control, DR and SCR cohorts.

|  | Control | DR | SCR | P values | | |
|---|---|---|---|---|---|---|
|  |  |  |  | Control vs. DR | Control vs. SCR | DR vs. SCR |
| **BVT (SCP)** | 1.11 ± 0.07 | 1.18 ± 0.04 | 1.25 ± 0.08 | 0.016 | <0.001 | 0.022 |
| **BVC (µm)(SCP)** | 17.47 ± 1.9 | 20.32 ± 3.8 | 21.43 ± 2.4 | 0.024 | 0.019 | 0.836 |
| **VPI (SCP)** | 10.26 ± 1.32 | 8.76 ± 2.60 | 8.43 ± 0.79 | 0.012 | 0.036 | 0.325 |
| **BVD (%)** |  |  |  |  |  |  |
| **C1 (SCP), 2mm** | 40.16 ± 10.32 | 35.61 ± 8.11 | 36.08 ± 7.02 | 0.019 | 0.051 | 0.154 |
| **C2 (SCP), 4mm** | 47.53 ± 6.32 | 40.72 ± 4.17 | 44.24 ± 5.52 | <0.001 | <0.001 | 0.017 |
| **C3 (SCP), 6mm** | 47.96 ± 2.36 | 38.89 ± 3.31 | 42.84 ± 3.24 | <0.001 | <0.001 | 0.014 |
| **C1 (DCP), 2mm** | 42.72 ± 13.19 | 38.98 ± 6.09 | 40.28 ± 10.17 | 0.024 | 0.058 | 0.208 |
| **C2 (DCP), 4mm** | 49.16 ± 5.78 | 43.32 ± 7.09 | 42.20 ± 4.17 | <0.001 | 0.011 | 0.095 |
| **C3 (DCP), 6mm** | 48.97 ± 3.18 | 41.75 ± 6.53 | 43.29 ± 4.30 | <0.001 | <0.001 | 0.005 |
| **FAZ-A (SCP), mm²** | 0.30 ± 0.06 | 0.39 ± 0.04 | 0.43 ± 0.05 | 0.006 | <0.001 | 0.008 |
| **FAZ-A (DCP), mm²** | 0.39 ± 0.08 | 0.52 ± 0.06 | 0.54 ± 0.06 | 0.006 | 0.005 | 0.059 |
| **FAZ-CI (SCP)** | 1.14 ± 0.11 | 1.38 ± 0.13 | 1.47 ± 0.14 | <0.001 | <0.001 | 0.004 |
| **FAZ-CI (DCP)** | 1.18 ± 0.12 | 1.41 ± 0.10 | 1.53 ± 0.13 | <0.001 | <0.001 | 0.002 |

[a]All values are presented as mean ± SD.

Table S2. Univariate analysis of individual OCTA features for NPDR stages.

| | Mild NPDR | Moderate NPDR | Severe NPDR | P values | | |
|---|---|---|---|---|---|---|
| | | | | Mild vs. Moderate | Moderate vs. Severe | Severe vs. Mild |
| **BVT (SCP)** | 1.14 ± 0.05 | 1.17 ± 0.06 | 1.23± 0.04 | 0.260 | 0.546 | 0.017 |
| **BVC (µm)(SCP)** | 18.06± 1.9 | 21.04 ± 2.2 | 21.86 ± 1.7 | 0.036 | 0.213 | 0.011 |
| **VPI (SCP)** | 9.94 ± 0.38 | 8.56 ± 0.15 | 7.79 ± 0.21 | 0.025 | 0.044 | <0.001 |
| **BVD (%)** | | | | | | |
| **C1 (SCP), 2mm** | 36.62± 9.03 | 36.01 ± 5.81 | 34.20 ± 9.38 | 0.019 | 0.154 | 0.041 |
| **C2 (SCP), 4mm** | 44.36 ± 6.72 | 40.81 ± 5.22 | 36.98 ± 6.50 | <0.001 | <0.001 | 0.005 |
| **C3 (SCP), 6mm** | 43.85 ± 3.38 | 38.95 ± 4.65 | 33.87 ± 4.24 | <0.001 | <0.001 | 0.014 |
| **C1 (DCP), 2mm** | 40.88 ± 10.37 | 38.78 ± 7.01 | 37.29 ± 8.16 | 0.042 | 0.658 | 0.018 |
| **C2 (DCP), 4mm** | 47.42 ± 4.83 | 43.39 ± 6.39 | 39.16 ± 7.25 | <0.001 | 0.026 | <0.001 |
| **C3 (DCP), 6mm** | 41.75 ± 11.08 | 42.32 ± 7.45 | 37.73 ± 5.29 | <0.001 | 0.006 | <0.001 |
| **FAZ-A (SCP), mm$^2$** | 0.33 ± 0.05 | 0.38 ± 0.07 | 0.46 ± 0.06 | <0.001 | <0.001 | <0.001 |
| **FAZ-A (DCP), mm$^2$** | 0.46 ± 0.07 | 0.53 ± 0.12 | 0.58 ± 0.09 | 0.018 | 0.003 | <0.001 |
| **FAZ-CI (SCP)** | 1.29 ± 0.14 | 1.38 ± 0.14 | 1.46 ± 0.18 | <0.001 | 0.002 | <0.001 |
| **FAZ-CI (DCP)** | 1.31 ± 0.21 | 1.42 ± 0.19 | 1.49 ± 0.17 | <0.001 | 0.009 | 0.002 |

[a]All values are presented as mean ± SD

Table S3. Univariate analysis of individual OCTA features for SCR stages.

|  | Mild SCR | Severe SCR | P values (Mild vs. severe) |
|---|---|---|---|
| **BVT (SCP)** | 1.22 ± 0.07 | 1.28± 0.05 | <0.001 |
| **BVC (µm)(SCP)** | 18.82 ± 3.1 | 24.05 ± 2.6 | 0.385 |
| **VPI (SCP)** | 9.21 ± 0.26 | 9.64 ± 0.29 | 0.521 |
| **BVD (%)** |  |  |  |
| C1 (SCP), 2mm | 36.99 ± 6.13 | 35.16 ± 8.08 | 0.163 |
| C2 (SCP), 4mm | 46.35 ± 4.53 | 42.13 ± 8.29 | 0.097 |
| C3 (SCP), 6mm | 46.85 ± 6.29 | 38.83 ± 3.23 | 0.018 |
| C1 (DCP), 2mm | 41.88 ± 10.85 | 38.68 ± 11.26 | 0.364 |
| C2 (DCP), 4mm | 47.06 ± 7.89 | 37.4 ± 8.36 | 0.073 |
| C3 (DCP), 6mm | 46.05 ± 6.25 | 40.5 ± 6.23 | 0.004 |
| **FAZ-A (SCP), mm²** | 0.41 ± 0.19 | 0..45 ± 0.12 | <0.001 |
| **FAZ-A (DCP), mm²** | 0.52 ± 0.19 | 0.56 ± 0.17 | <0.001 |
| **FAZ-CI (SCP)** | 1.45 ± 0.12 | 1.50 ± 0.15 | <0.001 |
| **FAZ-CI (DCP)** | 1.50 ± 0.14 | 1.56 ± 0.16 | 0.002 |

[a]All values are presented as mean ± SD

## Classification performance

Table S4. Performance evaluation of multi-task classification algorithm using optimal feature combination.

| Parameters | Classification performance | | |
|---|---|---|---|
|  | Area under the ROC curve, AUC | Sensitivity (%) | Specificity (%) |
| **Control vs. Disease** | 0.98 | 97.84 | 96.88 |
| **DR vs. SCR** | 0.94 | 95.01 | 92.25 |
| **NPDR Staging** | 0.96 | 92.18 | 86.43 |
| **SCR Staging** | 0.97 | 93.19 | 91.60 |